\documentstyle[aps,prl,psfig]{revtex}

\draft
\begin{document}
\newcommand {\be}{\begin{equation}}
\newcommand {\ee}{\end{equation}}
\newcommand {\bea}{\begin{eqnarray}}
\newcommand {\eea}{\end{eqnarray}}
\newcommand {\nn}{\nonumber}

\title{Possible Fulde-Ferrell-Larkin-Ovchinnikov State in CeCoIn$_5$}

\author{Hyekyung Won$^1$, Kazumi Maki$^2$, Stephan Haas$^2$, 
Niels Oeschler$^3$, Franziska Weickert$^3$,
and Philipp Gegenwart$^3$}
\address{
$^1$ Department of Physics, Hallym University, Chunchon 200-702,
South Korea\\
$^2$ Department of Physics and Astronomy, University of Southern
California, Los Angeles, CA 90089-0484\\
$^3$ Max Planck Institute of Chemical Physics of Solids, D-01187 Dresden, 
Germany
}

\date{\today}
\maketitle

\begin{abstract}

Angular dependent magnetothermal conductivity experiments  
on CeCoIn$_5$ indicate that this compound
is a d$_{x^2-y^2}$-wave superconductor.
In this study, the low-temperature behavior of the upper
critical field is
measured in a single crystal of CeCoIn$_5$ along the directions
$\vec{H} \parallel \vec{a}$ and $\vec{H} \parallel \vec{c}$.
The data is compared with model calculations of the 
upper critical field in a d$_{x^2-y^2}$-wave superconductor.
It is found that 
the observed $H_{c2}(T)$ along $\vec{H} \parallel \vec{a}$
is consistent with a Fulde-Ferrell-Larkin-Ovchinnikov (FFLO)
state at low temperatures, $T <$ 0.7K, 
whereas for $\vec{H} \parallel \vec{c}$ the 
FFLO state appears to be 
absent in CeCoIn$_5$. 
Furthermore, it is predicted that the quasiparticle density of states 
in the FFLO state exhibits a complex peak structure which should be 
observable by scanning tunneling microscopy.

\end{abstract}

\vspace{0.5cm}
\noindent
{\bf I. Introduction}
\vspace{0.5cm}

Recent measurements on CeCoIn$_5$ have led to a renewed discussion 
of a possible high-field Fulde-Ferrell-Larkin-Ovchinnikov (FFLO) state
in unconventional superconductors.\cite{fulde,larkin}
In this state, the coupling of the magnetic field to the 
quasiparticle spins dominates over the orbital coupling, 
leading to pairing between exchange-split Fermi surfaces, and hence 
to a spatially non-uniform superconducting order parameter. For 
conventional superconductors, its realization appears to be practically
impossible because of two reasons.
First, the sample quality has to be in the superclean 
limit, i.e. the quasiparticle mean path needs to be much larger than 
the coherence length. Second, the Ginzburg-Landau parameter
$\kappa$, measuring the
ratio of the magnetic penetration depth vs. the superconducting 
coherence length, should be very large, i.e.  $\kappa \gg 10$.  

The recent synthesis of quasi-two-dimensional nodal superconductors,
such as the high-$T_c$ cuprates, the $\kappa$-(ET)$_2$ salts, and 
CeCoIn$_5$ has changed this situation dramatically. It appears that 
the above two conditions can be met in high-quality single crystal 
samples of these compounds. 
\cite{maki1,yang,shimahara,maki2}
These systems are quasi-two-dimensional, leading to a large 
Ginzburg-Landau parameter in a planar magnetic field.
Furthermore, unlike in the
conventional s-wave superconductors, the
stability region of the FFLO state is much more extended in
d$_{x^2-y^2}$-wave superconductors compared to conventional ones.
\cite{maki1,yang}

Indications for possible FFLO states in organic superconductors were 
already reported in 
$\lambda$-(BEDTS)$_2$GaCl$_4$ and
$\kappa$-(BEDT-TFF)$_2$Cu(NCS)$_2$.\cite{tanatar,singleton} 
In the first compound, a kink in the thermal conductivity points to 
a transition from a FFLO state to a vortex lattice. In the 
second material a similar feature in the magnetization was identified.  
Moreover, recent evidence for
d$_{x^2-y^2}$-wave order parameter symmetry
was found in $\kappa$-(BEDT-TFF)$_2$Cu(NCS)$_2$
by angular dependent magnetothermal conductivity measurements in a 
rotating magnetic field within the conducting crystal plane.\cite{izawa1,won1}
Moreover, it was 
observed that the upper critical field $H_{c2}(T)$ in the 
FFLO regimes decreases quasi-linearly
with temperature, in contrast to the rather weak temperature 
dependence of $H_{c2}$ in the absence of a FFLO state as $T\rightarrow 0$. 

More recently, a new heavy fermion compound, CeCoIn$_5$, was discovered.
This material superconducts below a critical temperature $T_c$ = 2.3K,
\cite{petrovic} and it has a layered structure similar 
to the high-$T_c$ cuprates. Angular dependent magnetothermal conductivity 
experiments indicate d$_{x^2-y^2}$-wave superconductivity 
in this material.\cite{izawa2} Furthermore, the temperature dependence 
of the upper critical field $H_{c2}(T)$ for both $\vec{H} \parallel \vec{a}$
and $\vec{H} \parallel \vec{c}$ was measured in single crystals of
CeCoIn$_5$.\cite{tayama}
It was observed that $H_{c2}(T)$ for $\vec{H} \parallel \vec{a}$
exhibits a quasi-linear temperature dependence in the proposed FFLO
regime, T $<$ 0.7K. 

In this paper, the upper critical field $H_{c2}(T)$
in a single crystal sample of CeCoIn$_5$ is determined from
thermal expansion and magnetorestriction measurements 
with field orientations $\vec{H} \parallel \vec{a}$
and $\vec{H} \parallel \vec{c}$.
A d$_{x^2-y^2}$-wave model calculation is used to explain 
the temperature dependence of
$H_{c2}(T)$ along both directions,
considering the orbital effect and 
the Pauli term. In particular, the possibility of a FFLO state is addressed.
For $\vec{H} \parallel \vec{c}$ it is found that the temperature dependence of 
$H_{c2}(T)$ can be fitted consistently to the experimental data without 
invoking a FFLO state. On the other hand, for $\vec{H} \parallel \vec{a}$
we observe that the inclusion of a $\vec{v} \cdot \vec{q}$ term arising 
from a FFLO state is crucial for a consistent description of the 
observed $H_{c2}(T)$ below T = 0.7K. In particular, the quasi-linear 
T-dependence of $H_{c2}(T)$ at low temperatures can be understood within 
this framework, which provides compelling 
evidence for a FFLO state in CeCoIn$_5$. In order to help to further 
test and scrutinize this model, we also determine the corresponding
quasiparticle density of states which should be
accessible to scanning tunneling
microscopy experiments.

\vspace{0.5cm}
\noindent
{\bf II. $H_{c2}(T)$ for $\vec{H} \parallel \vec{c}$}
\vspace{0.5cm}

The temperature dependence of the
upper critical field along the crystal c-direction
can be obtained from two coupled integral equations\cite{won2}
\bea
- \ln t & = & \int_0^{\infty} \frac{d u}{{\rm sinh} u} \left[
1 - \exp (- \rho u^2 ) \cos (h u) (1 + 2 \rho^2 u^4 C) \right], \\
- C \ln t & = & \int_0^{\infty} \frac{d u}{{\rm  sinh} u} \left\{
C - \exp (- \rho u^2 ) \cos (h u) \left[ \frac{\rho u^2}{12} 
+ C \left( 1 - 8 \rho u^2 + 12 \rho^2 u^4 - \frac{\rho^3 u^6}{3} \right)
\right] \right\},
\eea
where $t \equiv T/T_c$, $\rho \equiv (v^2 e H)/(8 \pi^2 T^2)$,
and $h \equiv (g \mu_B H)/(2 \pi T)$. Here it is assumed that the wave
function of the Abrikosov state for a d$_{x^2-y^2}$-wave superconductor
can be written as
\bea
| \Psi \rangle = \left( 1 + C \left( a^{\dagger} \right)^4 \right) 
| 0 \rangle,
\eea
where the ``vacuum" $| 0 \rangle$ is the Abrikosov state of an s-wave 
superconductor, and $a^{\dagger}$ is the raising operator of the Landau 
level. In other words, $| 0 \rangle$ is a combination of the N = 0 
Landau states. For d$_{x^2-y^2}$-wave superconductors 
an admixture of higher Landau states that are allowed by symmetry needs
to be included in order to account for structural changes in the vortex 
lattice.\\ 

\vspace{-1.9cm}
\begin{figure}[h]
\centerline{\psfig{figure=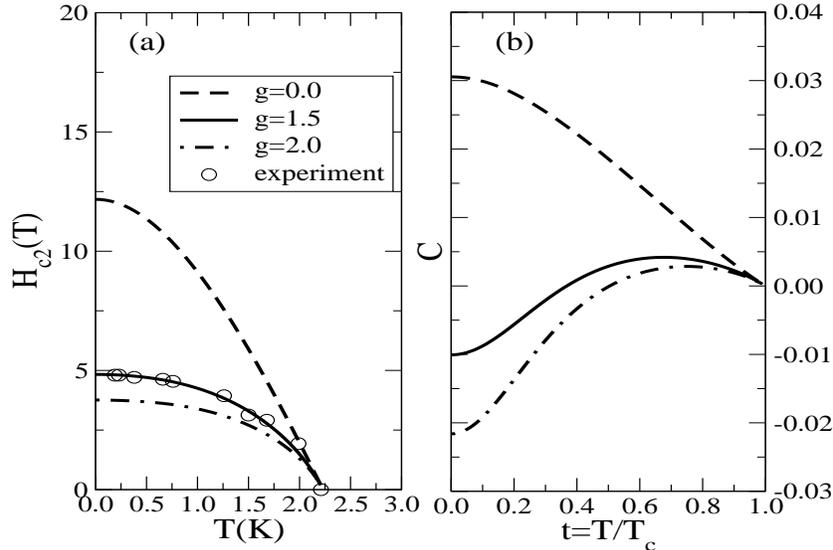,width=11cm,height=9cm,angle=0}}
\vspace{-0.5cm}
\caption{
Temperature dependence of (a) the upper critical field and (b) the 
admixture parameter $C$ in a 
d$_{x^2-y^2}$-wave superconductor with g-factors g=0, 1.5, and 2.
The magnetic field is applied along the crystal c-direction.
}
\end{figure}

In Fig. 1(a), experimental data of $H_{c2}(T)$ along the c-axis in CeCoIn$_5$ 
is compared with the numerical solution 
of Eqs. 1 and 2, describing a d$_{x^2-y^2}$-wave vortex lattice.
The temperature dependence of the upper critical field $H_{c2}(T)$ is 
extracted from low-temperature thermal expansion, $\Delta l(T,
B=\mbox{const})$, and magnetostriction $\Delta l(B,T=\mbox{const})$
measurements utilizing a high-resolution capacitive dilatometer at
temperatures down to 15\,mK and in magnetic fields up to 18\,T.
\cite{oeschler}
For
temperatures above $T_0\approx0.7\,\mbox{K}$, the superconducting-to-normal
phase transition is of second order and the $H_{c2}(T)$ is determined
from the midpoint of idealized jumps in $\partial\Delta l/\partial T$ and
$\partial\Delta l/\partial B$. Below $T_0$ sharp jumps in $\Delta l$ are
observed upon crossing $H_{c2}(T)$, indicative of a first order phase
transition\cite{bianchi1}.
The best fit with a numerical solution 
of Eqs. 1 and 2 is obtained with $v= 3.2738 \times 10^8$ cm/sec,
and $g$=1.5. For 
comparison, solutions with the same Fermi velocity $v$, but $g$= 0 and 2 
are also shown. The fit to the experiment appears to be very good in 
the low-temperature regime $T <$0.7K. Therefore, there are presently no obvious
indications for a FFLO state in this material 
by measurements of $H_{c2}(T)$ along the 
crystal c-direction.

In Fig. 1(b) we show the numerical solution for the admixture parameter $C$. 
Interestingly, for g$ \ge 1.2$, $C$ changes its sign as the temperature is decreased. 
Consequently, for g$=1.5$ one finds that the conventional hexagonal vortex 
lattice, which is stable at high temperatures, may change into a square vortex
for $T/T_c < 0.3$. This transition should be observable by small angle neutron 
scattering (SANS) with $\vec{H} \parallel \vec{c}$. A similar instability to a 
square vortex lattice was previously predicted for the high-$T_c$ cuprates 
in the vicinity of $H_{c2}$.\cite{won2}
Both, SANS\cite{keimer} and STM\cite{maggio}
on the vortex state of YBCO single crystals indeed indicate significant 
deviations
from a hexagonal towards a square vortex lattice in this compound. 
\cite{footnote1}
Furthermore, the above theory was recently extended to 
lower magnetic fields.\cite{shiraishi} 
The predicted square vortex lattice was observed by SANS
in a single crystal of LSCO at
$H$=2 Tesla.\cite{gilardi}

\vspace{0.5cm}
\noindent
{\bf III. $H_{c2}(T)$ for $\vec{H} \parallel \vec{a}$}
\vspace{0.5cm}

In order to match the experimental data for $H_{c2}(T)$ along the crystal 
a-axis, we explore the effect of
a $\vec{v} \cdot \vec{q}$ term arising from the 
formation of a FFLO state. 
\cite{won3} This leads to a new set of coupled integral equations,
\bea 
- \ln t & = & \int_0^{\infty} \frac{d u}{{\rm sinh} u} \left[
1 - \left\langle \exp (- \rho u^2 |s|^2 ) \cos \left[ h (1 - p \cos \phi ) u \right]
\left( 1 + \cos (4 \phi) \right) \left( 1 - 2 \rho u^2 s^2 C\right) 
\right\rangle \right] \\
- C \ln t & = & \int_0^{\infty} \frac{d u}{{\rm sinh} u} \left[
C - \left\langle \exp (- \rho u^2 |s|^2 ) \cos \left[ h (1 - p \cos \phi ) u \right]
\left( 1 + \cos (4 \phi) \right) \left( \rho u^2 s^2 + C \left( 1 - 4 \rho u^2 |s|^2
+ 2 \rho^2 u^4 |s|^4 \right) \right) \right\rangle \right],
\eea
where $s \equiv \sin \chi + i \sin \phi $, $\rho \equiv (v v_c e H)/(8 \pi^2 T^2)$,
$p \cos \phi \equiv (\vec{v} \cdot \vec{q})/(2 h)$, $\chi \equiv c k_z$, and 
$\langle ... \rangle$ is the angular average over $\phi $ and $\chi $. Here 
$\sqrt{v v_c} = 1.63 \times 10^8$cm/s is used, and following Gruenberg
and Gunther\cite{gruenberg}, we chose $\vec{q} \parallel \vec{H} \parallel \vec{a}$.
In this configuration, the vortex state is represented by
\bea
| \Psi \rangle = \left( 1 + C \left( a^{\dagger} \right)^2 \right) 
| 0 \rangle,
\eea
where the vacuum $| 0 \rangle $ is again the Abrikosov state of a simple s-wave
superconductor\cite{gruenberg}, but mixing now occurs with the N=2 Landau level.

\vspace{-1.5cm}
\begin{figure}[h]
\centerline{\psfig{figure=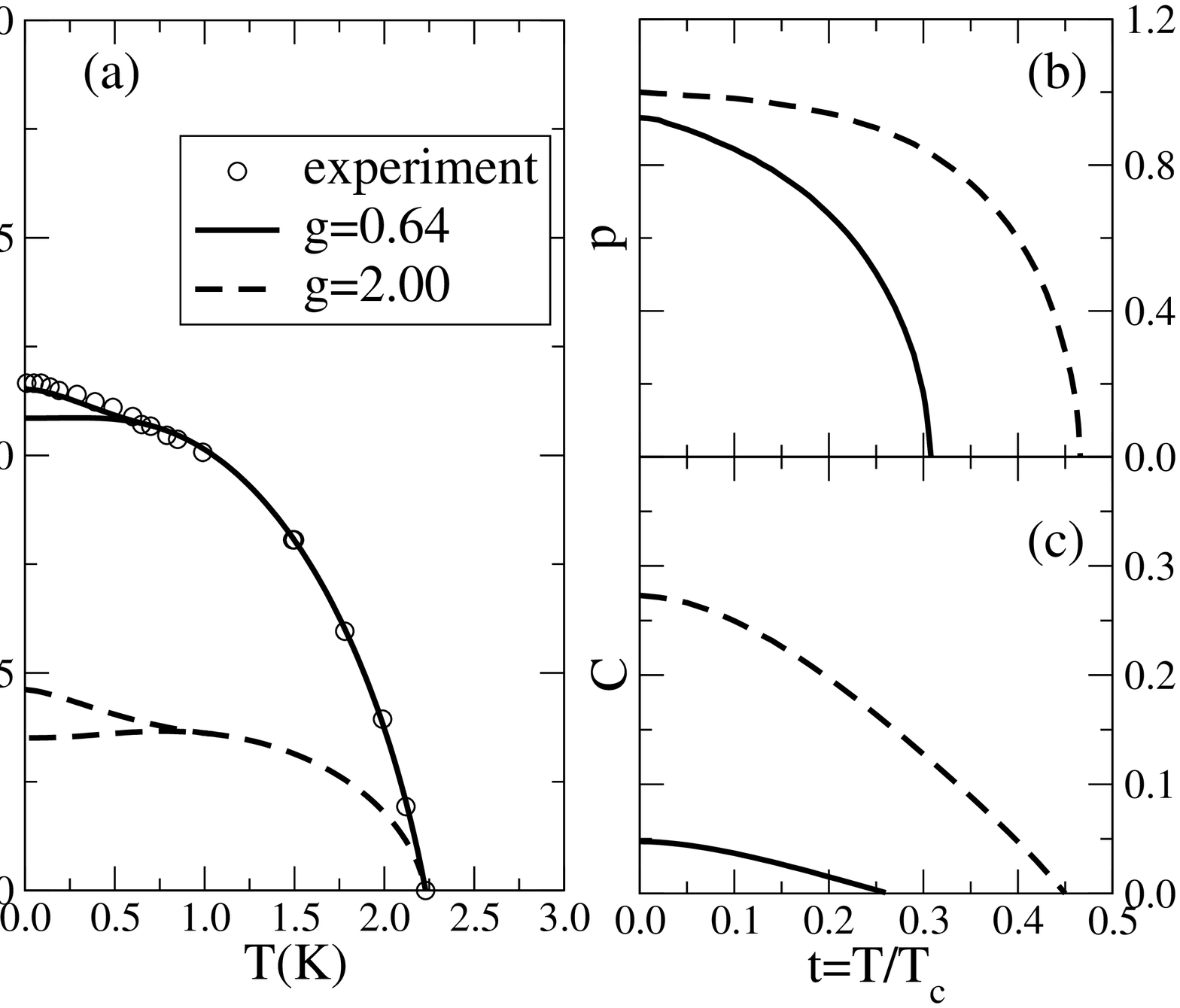,width=11cm,height=9cm,angle=0}}
\vspace{-0.5cm}
\caption{
Temperature dependence of (a) the upper critical field, (b) the $\vec{v} \cdot
\vec{q}/(2 H) = p \cos \phi$ term, and (c) the admixture parameter C in a  
d$_{x^2-y^2}$-wave superconductor with g-factors g= 0.64 (solid lines)
and 2 (dashed lines). In (a) the lower curves represent $p = 0$, i.e. absence 
of FFLO, whereas the upper curves have $p = 0.9$. 
The magnetic field is applied along the crystal a-direction.
The experimental data
(circles) is best described by g=0.64 and $p = 0.9$. 
}
\end{figure}

Furthermore, let us note that for the purpose of
In Fig. 2(a)
$H_{c2}(T)$ is shown for g= 0.64 and 2 along with the measurements of 
the upper critical field along the a-axis of CeCoIn$_5$.
We find that without the FFLO state ($p = 0$) one obtains a good
fit to the experiment down to
$T$ = 0.7K with g=0.64. However, for $T <$ 0.7K the measured upper critical 
field is approximately
linear in temperature. This feature can be reproduced by including a 
$\vec{v} \cdot
\vec{q}$ term due to the FFLO state with $p = 0.9$.
For comparison, we also show results for g = 2 which yield a zero-temperature 
critical field that is less than half the value detected in the experiment. 
In Fig. 2(b) $p(t)$ is shown for g = 0.64 and g = 2. 
From this plot it is clear that the FFLO region expands as g is increased. 
In Fig. 2(c) the coefficient
$C(t)$ is shown. Here we observe that $C$
exhibits a significant temperature dependence for g = 2, 
whereas for g = 0.64 the admixture
is almost negligible.\cite{footnote2}\\

Furthermore, let us note that for the purpose of 
the present discussion it was assumed that the transition at $H = H_{c2}$ is of
second order. However, a number of experiments on CeCoIn$_5$ 
indicate a possible first order transition, and onset
of magnetic order at $T \alt $ 0.8K.
\cite{izawa2,tayama,bianchi1,bianchi2} At the moment the nature of this
magnetic order is unknown. In case it is a spin density wave, the condensation 
energy is expected to be relatively small, and consequently 
its effect on $H_{c2}(T)$ 
should be small as well.\cite{yamaji} 

\vspace{0.5cm}
\noindent
{\bf IV. Quasiparticle Density of States}
\vspace{0.5cm}

Let us conclude this
discussion of a possible FFLO state in CeCoIn$_5$ by calculating 
the shape of the associated quasiparticle density of states.
In the vicinity of $H = H_{c2}$
and for $\vec{H} \parallel \vec{a}$ this observable is 
well approximated by\cite{maki2}
\bea
\frac{ N(E)}{N_0} - 1 = \frac{\Delta^2}{4\sqrt{\pi }} \sum_{\pm}
\left\langle 
\int_{- \infty }^{\infty } du \frac{ \exp (- u^2 ) \cos^2 (2 \phi )}
{\left[E \pm \tilde{H} ( 1 - p \cos \phi ) - \epsilon |s| u \right]^2}  
\right\rangle,
\eea 
where $\tilde{H} \equiv ( \mu_B g H)/2 $, $|s|= \sqrt{\sin^2(\phi) 
+ \sin^2(\chi)}$,
and $\epsilon \equiv \sqrt{v v_c e H}$.
Again, a finite $p$ indicates the presence of a FFLO state. 
This quasiparticle density of states is plotted in Fig. 3. For the parameters,
we have chosen $p = 0.9$ and $\epsilon / h = 0.2$, appropriate for
CeCoIn$_5$ in the temperature regime $T \alt 0.1K$.
In the absence of the FFLO state (Fig. 3(a)), there are two sharp 
resonances close
to $E = \pm H$. In the presence of the FFLO state, more structure appears
in the spectral response, as shown in Fig. 3 (b), with resonances at 
$E = \pm H (1 \pm p)$.
Therefore, precision measurements of the
quasiparticle density of states can provide a clear signal for the
presence of FFLO states and the symmetry of the underlying superconducting
order parameter. 
It should thus
be of interest to conduct a scanning tunneling microscope study of 
the quasiparticle density of states in 
CeCoIn$_5$ at $T <$ 0.7K in order to further scrutinize the proposed 
FFLO state.\\

\vspace{-4.0cm}
\begin{figure}[h]
\centerline{\psfig{figure=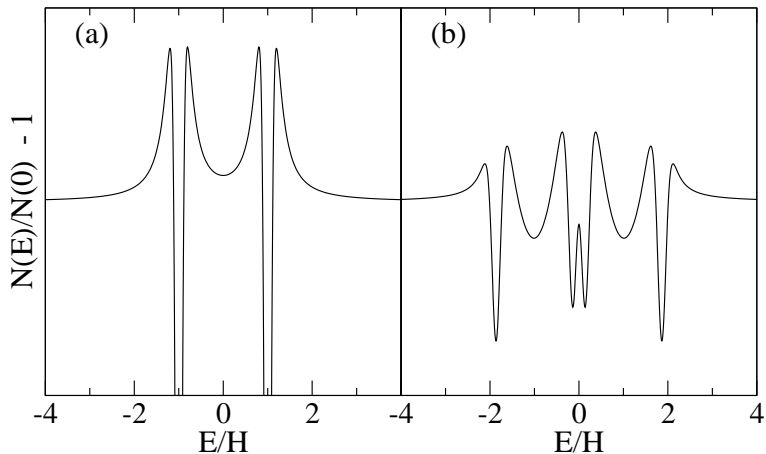,width=10cm}}
\vspace{-0.5cm}
\caption{Quasiparticle
density of states of a $d_{x^2 - y^2}$-wave superconductors in a 
magnetic field, (a)
in the absence of the FFLO state ($p = 0$), and (b) in the FFLO
state ($p = 0.9$). 
}
\end{figure}

\vspace{0.5cm}
\noindent
{\bf V. Conclusions}
\vspace{0.5cm}

In summary, the model calculation in this study incoporates consistently
(i) the d$_{x^2-y^2}$-wave symmetry of
the superconducting order parameter, (ii) the orbital effect, and (iii) 
a $\vec{v} \cdot \vec{q}$-term due to the formation of a FFLO state. The
model appears to describe well the observed temperature dependence 
of the upper critical field in CeCoIn$_5$. Furthermore, it indicates a 
significant renormalization of the g-factor in this compound, as well 
as the presence of a FFLO state at low temperatures if the applied field 
has an in-plane component. In this phase, 
the quasiparticle density of state 
is predicted to have a more complex structure. In order to further scrutinize 
the proposed model, it will be interesting to determine further 
relevant properties, such as the specific heat and the thermal conductivity.

\vspace{0.5cm}
\noindent
{\bf VI. Acknowledgements}
\vspace{0.5cm}

We thank J.L. Sarrao for providing a high-quality CeCoIn$_5$ single crystal.
Furthermore we are grateful to Y. Matsuda,
A. Ardavan, T. Ishiguro, R. Movshovich, M.-S. Nam, T. Roscilde,
and J. Sarrao for useful 
discussions on the FFLO state. S.H. acknowledges financial support by
the Petroleum Research Foundation and the National Science Foundation,
DMR-0089882.


\begin{references}
\bibitem{footnote1} While the observed apex angle of 
the vortex lattice is not 90$^o$ but rather 73$^o$-77$^o$, 
this deviation can be 
attributed to the orthorhombic structural distortion in YBCO. Moreover, the 
spatial anisotropy of the coherence length $\xi_a / \xi_b = 1.5$ is likely due 
to the presence of Cu-O chains in this compound. Precise STM measurements 
have revealed an
elliptical shape of the vortices, consistent with this ratio of 
$\xi_a / \xi_b$, and
\cite{maggio} the observed deviation of the apex angle from 90$^o$ is a likely 
consequence of this asymmetry.
\bibitem{footnote2} Furthermore, for larger $g$ the admixture coefficient
with the  N = 2 is greatly enhanced.
\end{references}
\end{document}